\documentstyle[12pt]{article}
\begin{document}
\thispagestyle{empty}
\noindent\hfill  OHSTPY-HEP-TH-93-014\\
\begin{center}\begin{Large}\begin{bf}
            Spontaneous symmetry breaking of (1+1)-dimensional
                $\phi^4$ theory in light-front field theory (II)\\
\end{bf}\end{Large}\vspace{.75cm}
 \vspace{0.5cm}
              Stephen S. Pinsky and Brett van de Sande\\[10pt]
\vspace{0.1cm}
                    {\em Department of Physics\\
                   The Ohio State University\\
                      174 West 18th Avenue\\
                       Columbus, OH~~43210}\\
\end{center}
\vspace{1cm}\baselineskip=35pt
\begin{abstract} \noindent
We discuss spontaneous symmetry breaking of (1+1)-dimensional
 $\phi^4$ theory
in light-front field theory using a Tamm-Dancoff truncation.
We show that, even though light-front field
theory has a simple vacuum state which is an eigenstate of the
full Hamiltonian, the field can develop a nonzero vacuum
expectation value.
This occurs because the zero mode of the field must
satisfy an operator valued constraint equation.
In the context of
(1+1)-dimensional
 $\phi^4$ theory we present solutions to the
constraint equation using a Tamm-Dancoff truncation
to a finite number of
particles and modes.
We study the behavior of the zero mode as a function
of coupling and Fock space truncation.
The zero mode introduces new
interactions into the Hamiltonian which breaks the $Z_2$ symmetry
of the theory when the
coupling is stronger than the critical coupling.
We investigate the energy spectrum in the symmetric and broken phases,
show that the theory does not break down in the vicinity of the critical
coupling, and discuss the connection to perturbation theory.
Finally, we study the spectrum of the field $\phi$ and
show that, in the broken phase, the field is localized away from
$\phi=0$ as one would expect from equal-time calculations.
We explicitly show that tunneling occurs.
\end{abstract}
\newpage\baselineskip=18pt
\section{Introduction}
Two fundamentally different
phenomenological pictures of hadronic matter have developed
over the past twenty years.
One is the constituent quark model where hadrons are
pictured as being made of a few heavy quarks bound by a
confining potential.  Here, Zweig's rule implies that very little
of a hadron's momentum is carried by gluons.
The other is the quark-parton model which is valid at
higher energy scales.  In this picture, a hadron is composed
of an indefinite number of light quarks and most of the
hadron momentum is carried by gluons.  We believe that
the underlying theory for both of these disparate pictures
is the non-abelian gauge theory
Quantum Chromodynamics (QCD).
It is believed that the light-front formulation of QCD is our
best hope of joining the constituent quark model with
the parton model and QCD.
In this elegant approach the field theory is
quantized on the null plane surface $x^+=(x^0+x^3)/\sqrt{2} = 0$
instead of the usual equal-time surface $x^0=0$.
This formulation avoids many of the
difficult problems that appear in the equal-time formulation
of field theory.

In the light-front  formalism the energy operator does not have a
square root operator, and, consequently, the vacuum
structure is relatively simple.
Dirac~\cite{dirac}, in 1949, showed that a maximum number of
Poincar\'e generators become independent of the dynamics
in the
``front form'' formulation, including certain Lorentz boosts.
The eigenstates of the light-front Hamiltonian
have Lorentz scalars $\hat{M}^2=P^2$ as eigenvalues
and describe bound states of arbitrary four-momentum and invariant
mass $\hat{M}$, allowing  easy computation of scattering amplitudes and
other dynamical quantities.
However, the most remarkable feature of this formalism
is the apparent simplicity of the vacuum.
The Fock space vacuum is an eigenstate of the full
Hamiltonian.
In other words, the bare vacuum {\em is}
the physical vacuum.
More detail is given in a recent review~\cite{PW}.

For the past several years an increasingly large group of
physicists has attempting to combine
Tamm-Dancoff~\cite{Tamm} procedures with light-front
quantization in order to perform
non-perturbative calculations in quantum field theory.
The assumptions, consistent with the constituent quark model picture,
are that a
hadron is well approximated by only a few particles
and that adding more particles
only refines this initial approximation.
This is in stark contrast to the
equal-time formulation of QCD where an infinite number
of gluons are essential to
construct even the vacuum~\cite{PW}.
If these efforts are successful, they could lead to
procedures for calculating not only the hadron mass spectrum but all the
quantities which depend on hadron wavefunctions:
structure functions,
fragmentation functions, {\em et cetera}.

The origin of this remarkably simple vacuum in the light-front formalism is
that
conservation of $P^+$ requires $\sum_{i} k^+_i =0$.
However, $k^+_i > 0$ for massive
particles, and the total light-front
Hamiltonian  annihilates the perturbative vacuum.
In contrast, the physical vacuum in equal-time quantization
is a highly complex composite of pair fluctuations
which is believed
to produce all phenomena that require long range order:
color confinement, chiral symmetry breaking, the
Goldstone pion, {\em et cetera}.
How can one obtain such
non-trivial  properties in the light-front
formulation of field theory?
These phenomena occur in light-front field theory because
the field includes a zero mode operator~\cite{maskawa} which is not
an independent degree of freedom.
This mode is a complicated operator-valued
function of all other modes in the theory and, since it carries zero
momentum,
 it can provide long range order.

This problem has recently been attacked from several directions.
The question of whether boundary conditions can be consistently defined
in light-front quantization has been discussed by
McCartor and Robertson~\cite{mccartor}  and Lenz~\cite{Lenz}.
They have shown that for massive theories
the energy and momentum derived from light-front
quantization are conserved
and are equivalent to the energy and momentum
one would normally write down in an equal-time theory.
In the analyses of Lenz {\em et al.}~\cite{Lenz} and
Hornbostel~\cite{hornbovac} one traces the fate of the equal
time vacuum in the limit  $P^3 \to \infty$ and equivalently
in the limit $\theta \to \pi/2$ when
rotating the evolution parameter $\tau= x^0 \cos \theta +
x^3 \sin \theta$ from the equal-time frame to
the light-front frame.
Heinzl {\em et al.}~\cite{heinzl} have considered
$\phi^4$ theory in (1+1)-dimensions and solved the zero-mode
constraint equation by truncating the equation to one
particle and retaining all modes.
They implicitly retain a two particle contribution
in order to obtain finite results. (We will discuss this
in detail later.)\@
They find a critical coupling for the onset of spontaneous
symmetry breaking.
Other authors~\cite{hari,robertson} find that, for  theories
allowing spontaneous symmetry breaking, there is a degeneracy
of light-front vacua and the true vacuum state
can differ from the perturbative vacuum through the addition of
zero mode quanta.
In addition to these approaches there are many
others~\cite{burkardt,hl,hcpauli}.

We present here a discussion of the zero-mode constraint
equation for (1+1)-dimensional $\phi^4$ field theory
[$\left(\phi^4\right)_{1+1}$] with symmetric
boundary conditions
and show how spontaneous
symmetry breaking occurs within the
context of this model.
This work builds upon the investigation of the one mode
case by Bender, Pinsky,
and van de Sande in Ref.~\cite{BPV}.
Our basic approach is to apply a Tamm-Dancoff truncation
to the Fock space.
We restrict the number of particles
and the number of modes to be less than some finite
limit.
This means that we can represent any operator by
a finite matrix and solve the operator
valued constraint equation
numerically.
The truncation assumes that states
with a large number of particles or large
momentum do not have an important
contribution to the zero mode.

We find the following general behavior:
for small coupling
(large $g$, where\linebreak[4]\ $g \propto 1/{\rm coupling}$)
the constraint
equation has a single solution and the field has no
vacuum expectation value (VEV).
As we increase the coupling (decrease $g$)
to what we will call the ``critical coupling''
$g_{\rm critical}$, two
additional solutions which give the
field a nonzero VEV appear.
These solutions differ
only infinitesimally from the first solution near the
critical coupling, indicating
the presence of a second order phase transition.
Above the critical coupling ($g < g_{\rm critical}$),
there are three solutions:  one with zero VEV,
the ``unbroken phase,'' and two with nonzero VEV,
the ``broken phase.''
As shown in Fig.~1, one
can plot the VEV as a function of $g$.  We will
call these the ``critical curves.''

In the weak coupling limit ($g$ large) the
solution to the constraint equation can
be obtained in perturbation theory.  This
solution does not break the $Z_2$ symmetry
and is believed to simply insert
the missing zero momentum contributions into
internal propogators.  This must
happen if light-front perturbation theory
is to agree with equal-time perturbation
theory~\cite{root}.  This has been shown for $\phi^3$ theory in
(1+1)-dimensions~\cite{maeno} and we believe
that it happens here as well.

Since the vacuum in this theory is trivial, all of the long range
properties
must occur in the operator structure of the Hamiltonian.
Above the critical coupling ($g < g_{\rm critical}$)
quantum oscillations spontaneously break the
$Z_2$ symmetry of the theory.
In a loose analogy with a symmetric
double well potential,
we have two new Hamiltonians for the
broken phase, each producing states localized in one of the wells.
The structure of the two Hamiltonians is determined from the
broken phase
solutions of the zero mode constraint equation.
We find that the two Hamiltonians have
equivalent spectra.
In a discrete theory without zero modes it is well known that, if one
increases
the coupling sufficiently, quantum correction will generate tachyons
causing the
theory to break down near the critical coupling.
Here the zero mode generates new interactions that prevent
tachyons from developing.   In effect what happens is that, while
quantum
corrections attempt to drive the mass negative, they also change the
vacuum energy
through the zero mode and the diving mass eigenvalue can never catch the
vacuum eigenvalue.
Thus, tachyons never appear in our spectra.

Another way to investigate the zero mode is to study the
spectrum of the field operator  $\phi$.
Here we find a picture that agrees with the symmetric double
well potential analogy.  In the broken phase, the field
is localized in one of the minima of the
potential and there is tunneling to the other minimum.

In Sec.~II we discuss the derivation of the
zero-mode constraint equation and the formal
operator constraint equation that
relates the zero mode to all the other modes in the problem.
In Sec.~III we apply a truncation allowing
only the first nonzero mode and many particles.
In Sec.~IV we examine the truncation where
we allow many modes and only one particle.
Next, in Sec.~V, truncations with several
particles and several modes are examined.
In Sec.~VI, we examine the spectrum of the field
operator $\sqrt{4 \pi} \phi$.
Finally, in Sec.~VII we discuss our results and the remaining work
needed on this problem.
\section{Quantization}
\setcounter{equation}{0}
The details of the Dirac-Bergmann prescription and its
application to the
system considered in this paper are discussed elsewhere
in the literature~\cite{heinzl,BPV,wittman}.
In this section, we will summarize those results and
introduce our notation.
We define light-front coordinates
$x^\pm = \left(x^0 \pm x^1\right)/\sqrt{2}$.
For a classical field the $\left(\phi^4\right)_{1+1}$
Lagrangian is
\begin{equation}
{\cal L} = \partial_+\phi\partial_-\phi - {{\mu^2}\over 2}
\phi^2 - {\lambda
\over 4!} \phi^4\;.
\end{equation}
We put the system in a box of length $d$ and impose
periodic boundary conditions.
For most of our discussion we work in momentum space.
We define $q_k$ by
\begin{equation}
\phi\!\left(x\right) = {1\over{\sqrt d}} \sum_n q_n(x^+)\,
e^{ik_n^+ x^-}\; ,
\end{equation}
where $k_n^+ =2\pi n/d$ and summations run over all integers
unless otherwise noted.
Following the Dirac-Bergmann prescription, we can identify
first-class constraints which define the conjugate momenta
and a secondary constraint
which determines the ``zero-mode'' $q_0$ in terms of the other
modes in the theory.
This result can also be obtained by integrating the
equations of motion
in position space or differentiating the Hamiltonian with
respect to the zero mode~\cite{BPV}.

Quantizing, we define creation and annihilation operators
$a_k^\dagger$ and $a_k$ by,
\begin{equation}
q_k = \sqrt {d\over{4 \pi \left| k \right|}} \: a_k\;,
\quad a_k = a_{-k}^\dagger\;,\quad  k\neq 0\; ,
\end{equation}
which satisfy the usual commutation relations
\begin{equation}
\left[a_k,a_l\right] =0\;,
\quad\left[a_k^\dagger,a_l^\dagger\right] =0\;,\quad
\left[a_k,a_l^\dagger\right] =\delta_{k, l}\; ,\quad k,l > 0\; .
\end{equation}
Likewise, we define the zero mode operator
\begin{equation}
q_0 = \sqrt{d\over{4 \pi}} \: a_0\;.
\end{equation}
It is useful to define the quantity,
\begin{equation}
\Sigma_n = \frac{1}{n!} \sum_{i_1, i_2, \ldots, i_n \neq 0}
\delta_{i_1+i_2+\cdots+i_n,0} \frac{: a_{i_1} a_{i_2} \cdots a_{i_n}:}{
\sqrt{\left| i_1 i_2 \cdots i_n\right|}} \; .
\end{equation}
General arguments suggest that the Hamiltonian
should be symmetric ordered~\cite{benderpinsky}.
However, it is not clear how one should treat the
zero mode since it is not a dynamical field.
As an {\em ansatz} we will treat $a_0$ as an ordinary
field operator when symmetric ordering the Hamiltonian.

Rescaling $P^-$, the quantum Hamiltonian is~\cite{BPV},
\begin{eqnarray}
\lefteqn{H ={{96 \pi^2}\over{\lambda d}} P^- =} \nonumber\\
& &\frac{g}{2} a_0^2 + \frac{a_0^4}{4}  + g \Sigma_2+ 6 \Sigma_4
  \nonumber\\
&+&\frac{1}{4}\sum_{n\neq 0} {1\over{|n|}} \left( a_0^2 a_n a_{-n}
+ a_n a_{-n}
a_0^2 + a_n a_0^2 a_{-n} \right. \nonumber\\
& & \quad \left. +  a_n a_0 a_{-n} a_0 + a_0 a_n a_0 a_{-n} +
a_0 a_n a_{-n} a_0 - 3 a_0^2\right) \nonumber\\
&+& \frac{1}{4}\,\sum_{k, l, m \neq 0}\, {\delta_{k+l+m, 0}
\over{\sqrt{\left|k l m \right|}}}
{\left(a_0 a_k a_l a_m + a_k a_0 a_l a_m + a_k a_l a_0 a_m +
a_k a_l a_m a_0 \right)}   \nonumber \\
&-&   C  \; .
\end{eqnarray}
where $g= 24 \pi \mu^2/\lambda$.
We have removed tadpoles
from the symmetric ordered Hamiltonian by
normal ordering the third and fourth terms and subtracting,
\begin{equation}
-{3\over{4}}\; a_0^2\sum_{n\neq 0} {1\over{|n|}} \; .
\end{equation}
In addition, we have subtracted a constant $C$ so that the
VEV of $H$ is zero.
Note that this renormalization prescription is equivalent
to a conventional mass renormalization and
does not introduce any new operators (aside from the constant)
into the Hamiltonian.
The constraint equation for the zero-mode can be obtained by taking a
derivative of $P^-$ with respect to $a_0$. Consequently, we
symmetric order the
constraint equation:
\begin{equation}
0 = g a_0 + a_0^3 +  \sum_{n\neq 0} {1\over{|n|}}
\left( a_0 a_n a_{-n} + a_n a_{-n} a_0 + a_n a_0 a_{-n} -
\frac{3 a_0}{2} \right)+ 6 \Sigma_3 \; . \label{constraint}
\end{equation}
Using the constraint equation, we can rewrite $H$ as:
\begin{eqnarray}
H &=& g \Sigma_2 + 6 \Sigma_4 -{a_0^4\over 4}
+\frac{1}{4}\sum_{n\neq 0} \frac{1}{|n|}
\left( a_n a_0^2 a_{-n} - a_0 a_n a_{-n} a_0\right) \nonumber \\
 & & +\frac{1}{4} \sum_{k, l, m \neq 0} \frac{\delta_{k+l+m, 0}}
{\sqrt{|k l m |}}
\left( a_k a_0 a_l a_m + a_k a_l a_0 a_m\right)  -  C \; .
 \label{hamiltonian}
\end{eqnarray}
It is clear from the general structure of (\ref{constraint})
that $a_0$ as a function of the other modes is not necessarily
odd under the transform $a_k \to -a_k$, $k \neq 0$
associated with the $Z_2$ symmetry of the system.
Consequently, the zero mode can induce $Z_2$ symmetry breaking
in the Hamiltonian.
Along with the zero mode and the Hamiltonian,
$\Sigma_n$ commutes with the longitudinal momentum
operator,
\begin{equation}
             \left[a_0,P^+\right] = 0 \quad
             \left[H,P^+\right] = 0 \quad
             \left[\Sigma_n,P^+\right] = 0  \; .
\end{equation}

In order to render the problem tractable, we impose a
Tamm-Dancoff truncation on the Fock space.
Define $M$ to be
the number of nonzero modes and $N$ to be the
maximum number of allowed particles.  Thus, each state in
the truncated Fock space can be represented by a vector of
length $S=\left(M+N\right)!/\left(M! N! \right)$ and  operators
can be represented by $S \times S$ matrices.
One can define the usual Fock space basis,
\begin{equation}
\left|n_1,n_2,\ldots,n_M\right\rangle =
\frac{\left(a_1^\dagger\right)^{n_1}}{\sqrt {n_1!}}
\frac{\left(a_2^\dagger\right)^{n_2}}{\sqrt {n_2!}}
\cdots
\frac{\left(a_M^\dagger\right)^{n_M}}{\sqrt {n_M!}}
\left| 0 \right\rangle
\end{equation}
where $n_1 + n_2 + \ldots +n_M \leq N$.
Thus, for example,
\begin{eqnarray}
P^+ \left|n_1,n_2,\ldots,n_M\right\rangle &=&
\frac{2 \pi}{d} \left(n_1 +2 n_2 + \cdots + M n_M \right)
\left|n_1,n_2,\ldots,n_M\right\rangle \\
\Sigma_2 \left|n_1,n_2,\ldots,n_M\right\rangle &=&
\left(\frac{n_1}{1} + \frac{n_2}{2} + \cdots + \frac{n_M}{M} \right)
\left|n_1,n_2,\ldots,n_M\right\rangle \; .
\end{eqnarray}
In matrix form, $a_0$ is real and symmetric. Moreover, it is
block diagonal in states of equal $P^+$ eigenvalue.
Due to the Fock space truncation, the commutator
$\left[a_k,a_l^\dagger\right]$ is somewhat different than
its full Fock space counterpart.  We will use the full Fock
space commutators to move creation operators to the left as
much as possible in a product before performing any matrix
multiplications.
\section{One Mode, Many Particles}
\setcounter{equation}{0}
In this section we review the results of Ref.~\cite{BPV}.  Consider
 the case of one mode $M=1$ and many particles.
In this case,
the zero-mode is diagonal and can be written as
\begin{equation}
a_0 = f_0 \left| 0 \right\rangle\left\langle 0 \right| + {\sum_{k=1}^N
f_{k} \left| k \right\rangle\left\langle k \right|}\; . \label{onemode}
\end{equation}
Note that $a_0$ in (\ref{onemode}) is even under
$a_k \to -a_k$, $k \neq 0$ and any non-zero solution
breaks the $Z_2$ symmetry of the original Hamiltonian.
The VEV is given by
\begin{equation}
\langle 0 | \phi | 0 \rangle ={1\over{\sqrt{4 \pi}}}
\langle 0| a_0 | 0\rangle = {1\over{\sqrt{4 \pi}}} f_0 \; .
\end{equation}
Substituting (\ref{onemode}) into the constraint equation
(\ref{constraint}) and sandwiching the
constraint equation between Fock states, we get a
recursion relation for $\left\{f_{n}\right\}$:
\begin{equation}
0 =  g f_{n} + {f_{n}}^{3} + (4n - 1) f_{n} +
\left(n+1\right) f_{n+1} + n f_{n-1} \label{recursion}
\end{equation}
where $n \leq N$, and we define $f_{N+1}$ to
be unknown.
Consequently,
$\left\{ f_{1}, f_{2}, \ldots, f_{N+1} \right\}$
is uniquely determined by a
given choice of $g$ and $f_0$.
In particular, if $f_{0}=0$ all the $f_{k}$'s are zero
independent of $g$.
This is the unbroken phase.

Our first objective is to determine which solutions
of the recursion relation are consistent with the
Tamm-Dancoff truncation.
If $f_{n}$ is large for large
$n$ then the high energy levels will be strongly affected.
The paradigm for spontaneous symmetry breaking is the
symmetric double well potential in ordinary quantum
mechanics which has a ground
state centered in either well rather than at the symmetry point.
This paradigm
indicates that the behavior of the system is unaffected by the barrier
for
energies far above the barrier separating the wells.
Hence, we only seek solutions where $f_{n}$ is small for large $n$.
We caution here that this paradigm does not translate entirely to the
light-front formulation since the symmetry breaking occurs in the
Hamiltonian
and the vacuum itself is unaffected.

We begin our analysis of Eq.~(\ref{recursion}) by
dropping the cubic term and finding
solutions to the resulting linear system.
Assume for now that $N=\infty$ and that
$\left|f_n\right|$ is decreasing with $n$.
For large $n$, the terms linear in $n$  dominate,
Eq.~(\ref{recursion}) becomes
\begin{equation}
f_{n+1} + 4 f_{n} + f_{n-1} = 0\; .
\end{equation}
There are two solutions to this equation:
\begin{equation}
f_{n} \propto \left(\sqrt{3} \pm 2\right)^{n}\; .
\end{equation}
We reject the plus solution because it grows with $n$.
Dropping the cubic
term from (\ref{recursion}) we define the generating function
\begin{equation}
F(z) = \sum_{n=0}^{\infty} f_{n} z^n\; .
\end{equation}
If $f_{n}$ goes like $\left( \sqrt{3} -2\right)^n$
then
the radius of convergence of $F(z)$ is $2+\sqrt{3}$ and we expect
$F(z)$ to be singular at $|z| = 2+\sqrt{3}$.
Similarly, if $f_{n} \sim\left(\sqrt{3} + 2\right)^n$,
then we expect $F(z)$
to be singular at $|z| = 2 -\sqrt{3}$.

The function $F(z)$ satisfies a differential
equation whose solution is
\begin{equation}
\frac{F(z)}{F(0)}=\left( {{z+2-\sqrt 3}\over {2
-\sqrt 3}}\right)^{-{{\sqrt 3 - 3 + g}\over{2\sqrt 3}}}
\left({{z+2+\sqrt 3}
\over{2 + \sqrt 3}}\right)^{- {{\sqrt 3 +3-g}\over{2\sqrt 3}}}\; .
\end{equation}
Note that this solution for $F(z)$ has
singularities at the expected values of $z$.
If we want $f_{n}$ to have the asymptotic behavior
$\left(\sqrt{3} - 2 \right)^n$ for large $n$,
then we must eliminate the branch
point of $F(z)$ at $|z| =2-\sqrt 3$.
This gives the condition
\begin{equation}
-{{\sqrt{3} - 3 + g}\over{2\sqrt 3}} = K\;,
\quad K= 0, 1, 2\ldots \label{criticals}
\end{equation}
Concentrating on the $K=0$ case, we find a critical coupling
\begin{equation}
g_{\rm critical} = 3 - \sqrt{3}
\end{equation}
or
\begin{equation}
\lambda_{\rm critical}=4\pi\left(3+\sqrt{3}\right) \mu^2 \approx 60\mu^2,
\end{equation}
consistent with equal-time calculations~\cite{chang}.

The solution to the linearized equation
is an approximate solution to the full equation~(\ref{recursion})
for $f_0$ sufficiently small.
We need to determine solutions of
the full nonlinear equation which
converge for large $n$.
We will use both the $\delta$-expansion and numerical
methods to do this.\\

The $\delta$-expansion is a powerful perturbative technique
for linearizing nonlinear problems.
It has been shown to be an accurate technique for solving
problems in differential equations, quantum mechanics,
and quantum field theory~\cite{bendermilton}.

We rewrite Eq.~(\ref{recursion}) as
\begin{equation}
\left(g - 1 + 4 n\right)f_{n} + f_{n}^{1+2\delta} + \left(n +
1\right) f_{n+1} + n f_{n-1} = 0\; .
\end{equation}
Setting $\delta = 0$ gives the linear finite difference equation which
is the zeroth-order approximation in the $\delta$-expansion. One then
expands in
powers of $\delta$ about $\delta = 0$.
One recovers the problem of interest
at $\delta = 1$.

We find to first order the critical curve for $\delta = 1$,
\begin{equation}
g = \left(2-\sqrt 3\right) \left(1+{1\over{\sqrt 3}} \ln \left(2+\sqrt
3\right)\right) - \ln {f_{0}^2}\; .
\end{equation}
These results are plotted as the dashed curve in Fig.~1.
The expansion behaves badly near $f_{0}=0$
because of the $\ln f_{0}^2$ term in the expansion.
The $\delta$-expansion
analysis clearly shows that there is a critical curve and
not merely a critical
point.\\

We can also study the critical curves by looking for
numerical solutions to Eq.~(\ref{recursion}). The method
used here is to find values of $f_{0}$ and $g$ such
that $f_{N+1}=0$.
Since we seek a
solution where $f_{n}$ is decreasing with $n$,
this is a good approximation.
We find that for $g>3-\sqrt 3$ the only real solution is $f_{n}=0$ for
all $n$.
For $g$ less than $3-\sqrt{3}$ there are two additional solutions.
Near
the critical point $\left| f_{0} \right|$ is small and
\begin{equation}
f_{n} \approx f_{0} \left(2 - \sqrt 3\right)^n\; .
\end{equation}
The critical curves are indicated by the solid lines in
Fig.~1.
These solutions converge quite rapidly with $N$.
The critical curve for the broken phase
is approximately parabolic in shape:
\begin{equation}
g \approx  3-\sqrt 3 - 0.9177 f_{0}^2 \; .
\end{equation}

It is instructive to study the behavior of the constraint
equation~(\ref{recursion}) away from the critical curves.
In Fig.~2 we plot $\left| f_{n} \right|$ as a function
of $n$ and $f_{0}$ for $g = 1.2$. We see that, as $n$ becomes large,
all the $\left| f_{n} \right|$ increase and as $f_{0}$
approaches the critical curve,
which is at $f_{0} \approx 0.2700$ for $g=1.2$,
all the $\left| f_{n}
\right|$'s decrease rapidly.
As $f_{0}$ increases beyond the critical curve the $\left|
f_{n}\right|$'s increase rapidly once again. The fact that
$\left| f_{n}\right|$
increases rapidly on both sides of the critical curve is a
manifestation of the nonlinearity in~(\ref{recursion}).

In Fig.~3 we extend the critical curves to the negative
$g$ region.  This region corresponds to negative
values of $\mu^2$ in the Hamiltonian.
We have not studied the spectrum for $\mu^2<0$ ($g<0$);
however, it is easy to see from the figures that nothing
unusual happens, at least for small $g$.
As predicted by Eq.~(\ref{criticals}),
there are additional solutions near the critical curve
for the unbroken phase.
The curves shown are independent of $N$ for $N$ large.
It is not clear how to interpret these solutions.

We can also study the eigenvalues of the Hamiltonian for the
one mode case.
The Hamiltonian is diagonal for this Fock space truncation
and,
\begin{equation}
\label{onehamiltonian}
\left\langle n \right| H \left| n \right\rangle =
{3\over 2} n (n-1) + n g
-{f_n^4\over 4} - {{2 n+1}\over 4} f_n^2 +{{n+1}\over 4}
f_{n+1}^2 + {n\over 4}
f_{n-1}^2 - C\; .
\end{equation}
The invarient mass eigenvalues are given by
\begin{equation}
   P^2 | n \rangle =
   2 P^+ P^- | n \rangle =
\frac{n \lambda  \langle n | H | n \rangle}{24 \pi} | n \rangle
\end{equation}
In Fig.~4 the dashed lines show
the first few eigenvalues as a function of $g$ without the zero-mode.
When we include the broken phase of the zero mode,
the energy levels shift as shown by the solid curves.
For $g < g_{\rm critical}$  the energy levels increase
above the value they had without the zero mode.
The higher levels change very little, as
our paradigm would suggest, because $f_n$ is small for large $n$.
\section{Many Modes, One Particle}
\setcounter{equation}{0}
Now, let us look at the case of one particle $N =1$ and many modes.
In this case, the zero mode is diagonal and can be written as
\begin{equation}
a_0 = b_0 \left| 0 \right\rangle\left\langle 0 \right| +
{\sum_{k=1}^M b_k \: a_k^\dagger \left| 0
\right\rangle\left\langle 0 \right| a_k }\; .
\label{oneparticle}
\end{equation}
Substituting this into the constraint equation (\ref{constraint})
and sandwiching the constraint equation between various states,
one obtains a system of equations for the coefficients
$\left\{b_k\right\}$:
\begin{eqnarray}
0 &=& b_0^3 + g b_0 + {\sum_{k=1}^M {{b_k-b_0}\over k}}
\label{beforemixing}\\
0 &=& b_m^3+g b_m +{4\over m} b_m + {b_0\over m} -
b_m {\sum_{k=1}^M {1\over k}} + {\sum_{k=1}^M {{\left\langle
0\right| a_m a_k a_0 a_k^\dagger a_m^\dagger \left| 0
\right\rangle}\over k}} \; . \label{mixing}
\end{eqnarray}
The last term of (\ref{mixing}) couples the one and
two particle sectors of $a_0$.
Assuming that the Fock space truncation is a good approximation,
we set this term equal to zero.
Note that $b_k =0$, $k=0,1,\ldots,M$, is always a solution of
this system of equations; this is the
unbroken phase.  Evaluating the system of equations
numerically, we find the critical curves shown in Fig.~5.

Let us calculate the critical coupling.
As with the one mode case, we drop the cubic terms
from (\ref{beforemixing}) and (\ref{mixing})
and find a solution to the resulting linear system.
Defining $\theta=g - \sum_{k=1}^M 1/k$, we find
\begin{eqnarray}
b_m = - \frac{b_0}{m \theta +4} \label{linconv}\\
\theta= \sum_{m=1}^M \frac{1}{m \left(m \theta+4\right)}
\end{eqnarray}
or, in the limit of large $M$,
\begin{equation}
4 \theta = \gamma_E + \Psi\!\left(1+4/\theta\right)+O(1/M)
\quad \theta = 0.6267537\ldots + O(1/M)
\end{equation}
where $\gamma_E$ is Euler's constant, and $\Psi\!\left(x\right)$
is the digamma function.
Thus, the critical coupling is logarithmically divergent,
\begin{equation}
\label{loggc}
g_{\rm critical}= \log M + \gamma_E + \theta + O(1/M) \; .
\end{equation}
We compare this expression with numerical results in Fig.~6.

Next we need to examine whether the Fock space truncation is
consistent, {\em i.~e.\ } whether $b_m$ converges for large $m$.
If we look at the linearized solutions, Eq.~(\ref{linconv}),
we see that $b_m$ does converge
near the critical coupling.
Now, examine the full non-linear equations.
{}From Eq.~(\ref{mixing}), we see that $b_m$ is independent of
$m$ if $b_m=-b_0/4$.
Substituting this into Eqs.~(\ref{beforemixing}) and (\ref{mixing}),
one can solve for $g$.
Define $g_0$ to be this value of $g$,
\begin{equation}
\label{g0}
g_0 = \frac{59}{60} \sum_{k=1}^M \frac{1}{k} \; .
\end{equation}
Note that $g_0 < g_{\rm critical}$ for
all $M$ and that $g_0$ is
logarithmically divergent in $M$.
For $g=g_0$, $b_m$ is independent of $m$.
For $g<g_0$, $b_m$ diverges with $m$,
and, for $g>g_0$, $b_m$ converges with $m$.
This means that the solution to the constraint
equation is inconsistent with the Fock space truncation
for $g\leq g_0$.

Elsewhere it has been suggested that one use a somewhat different
{\em ansatz} for the zero mode~\cite{heinzl},
\begin{equation}
\label{ansatz}
\tilde{a}_0 = c_0 + {\sum_{k=1}^M c_k a_k^\dagger a_k}\; .
\end{equation}
How does it compare with the definition in
Eq.~(\ref{oneparticle})?  Let us look at some matrix elements in each
case:
\begin{eqnarray}
\left\langle 0 \right| a_0 \left| 0 \right\rangle = b_0 &
\left\langle 0 \right| \tilde{a}_0 \left| 0 \right\rangle =
c_0\nonumber\\
\left\langle 0 \right| a_k a_0 a_k^\dagger\left| 0 \right\rangle =
b_k &
\left\langle 0 \right| a_k \tilde{a}_0 a_k^\dagger
\left| 0 \right\rangle = c_0 + c_k \nonumber\\
\left\langle 0 \right| a_k a_l a_0 a_l^\dagger a_k^\dagger
\left| 0 \right\rangle = 0 &
\left\langle 0 \right| a_k a_l \tilde{a}_0 a_l^\dagger
a_k^\dagger \left| 0 \right\rangle = c_0 + c_k +c_l\; , &
k\neq l \nonumber\\
\frac{1}{2}\left\langle 0 \right| \left(a_k\right)^2 a_0
\left(a_k^\dagger\right)^2 \left| 0 \right\rangle = 0 &
\frac{1}{2}\left\langle 0 \right| \left(a_k\right)^2
\tilde{a}_0 \left(a_k^\dagger\right)^2 \left| 0
\right\rangle = c_0 + 2 c_k \; .
\end{eqnarray}
The important point is that this {\em ansatz} makes very
different assumptions about the two particle sector.
If we identify $b_0$ with $c_0$ and $b_k$ with $c_0 + c_k$,
it would be equivalent to a different choice choice for the
last term in equation (\ref{mixing}):
\begin{equation}
{\sum_{k=1}^M {{\left\langle 0\right| a_m a_k a_0
a_{k}^\dagger a_m^\dagger \left| 0 \right\rangle}\over k}} =
{\sum_{k=1}^M {{b_k - b_0}\over k}}+b_m {\sum_{k=1}^M {1\over k}} +
{{2 b_m-b_0}\over m} \; .
\end{equation}
Using this assumption for the two particle sector gives us a
different system of equations for the coefficients,
\begin{eqnarray}
0 &=& b_0^3 + g b_0 + {\sum_{k=1}^M {{b_k-b_0}\over k}} \\
0 &=& b_m^3+g b_m +{6\over m} b_m  +
      {\sum_{k=1}^M {{b_k-b_0}\over k}} \; .
\end{eqnarray}
These are equivalent to equations (3.6a)  and (3.6b) in Heinzl,
{\em et~al.}~\cite{heinzl}.
In this case, the zero mode matrix elements along with the
critical coupling converge in the limit of large $M$.

How do we interpret this {\em ansatz} in terms of a Fock
space truncation?
We have assumed, in essence, that the zero mode
has a contribution from the
2~particle sector that cannot be neglected.
Is this the correct way to truncate the Fock space?
We will answer this question in Sec.~V.

We turn our attention to the Hamiltonian.
For this Fock space truncation the Hamiltonian is diagonal.
Substituting
Eq.~(\ref{oneparticle}) into Eq.~(\ref{hamiltonian}) and
sandwiching between states, we obtain
\begin{eqnarray}
\langle 0 | H | 0 \rangle&=&
\frac{1}{4}\sum_{k=1}^{M} \frac{b_k^2 -b_0^2}{k}
-\frac{b_0^4}{4}-C\\
\langle 0 | a_m H a_m^\dagger | 0 \rangle&=&
\frac{g}{m} -\frac{b_m^4}{4} - \frac{b_m^2}{2 m} -b_m^2 \sum_{k=1}^{M}
\frac{1}{k} +\frac{b_0^2}{4 m}-C \; . \label{mham}
\end{eqnarray}
We choose $C$ such that the VEV of $H$ is zero and
plot a spectrum in Fig.~7.  Note the vertical line
representing the value of $g_0$.  Presumably, to the left
of this line, the Fock space truncation breaks down.
Also, note that states with larger energy have smaller
longitudinal momentum $P^+$.
Clearly, the zero mode has a large effect on the spectrum.
\section{Many Modes and Many Particles}
\setcounter{equation}{0}
We turn our attention to the more general case
of many modes and many particles.
Many of the features that
were seen in the one mode and one particle cases
remain.
Plotting the VEV vs.\ $g$, one finds that
the critical curves have the same form with an
unbroken phase, a broken phase, and a critical coupling.
There is rapid convergence as one
increases $N$ and a logarithmic divergence as
one increases $M$.
However, now one can have more than one state of a given
momentum.  Thus, $H$ and $a_0$ are no
longer strictly diagonal and $\Sigma_3$ is nonzero.

Consequently, the zero mode for the unbroken phase is
no longer zero.
We can use Eq.~(\ref{constraint}) and solve
for the unbroken phase of $a_0$
perturbatively in $\lambda$ or, equivalently, in $1/g$:
\begin{equation}
\label{perturb}
a_0 = - \frac{6}{g} \Sigma_3 +
\frac{6}{g^2}\left(2 \Sigma_2 \Sigma_3+ 2 \Sigma_3
\Sigma_2+\sum_{k=1}^M \frac{a_k \Sigma_3 a_k^\dagger +
a_k^\dagger \Sigma_3 a_k -\Sigma_3}{k}\right)
+ O(1/g^3) \; .
\end{equation}
For each order in $1/g$, one can see that
$a_0$ is odd under the transform $a_k \to -a_k$, $k \neq 0$.
Consequently,
\begin{equation}
\langle \alpha | a_0 |\beta\rangle=0\, \mbox{
if }\,\langle \alpha| \hat{N}|\alpha \rangle -
\langle \beta |\hat{N}|\beta \rangle \mbox{ is even}
\end{equation}
where $\hat{N}$ is the number operator and
$|\alpha\rangle$ and $|\beta\rangle$ are
Fock space basis states.
In particular, the diagonal
matrix elements of $a_0$ are zero.
It is important to note that, as expected, this expansion does not
produce the broken phase~\cite{robertson}.
When substituted into the Hamiltonian,
the perturbative expansion of the
zero mode produces new interactions in perturbation theory.
It is generally believed that equal-time and light-front
perturbation theories are equivalent~\cite{root}
and that the zero modes will change the internal
propagator to be the full propagator that one
would have if there were a dynamical zero mode.
Maeno has, in fact, shown that this is true for
$\phi^3$ theory in (1+1)-dimensions~\cite{maeno}.

In order to calculate the zero mode for a given
value of $g$ one converts the constraint equation
(\ref{constraint}) into an $S \times S$ matrix equation
in the truncated Fock space.
This becomes a set of $S^2$ coupled cubic equations
and one can solve for the matrix elements of $a_0$
numerically~\cite{mma}.
Considerable simplification occurs because $a_0$ is
symmetric and is block diagonal in states of equal
momentum.
For example, in the case $M=3$, $N=3$,
the number of coupled equations is 34 instead
of $S^2=400$.
In order to find the critical
coupling, we take $\langle 0|a_0|0\rangle$ as given
and $g$ as unknown and solve the constraint
equation for $g$ and the other
matrix elements of $a_0$ in the limit of small
but nonzero $\langle 0|a_0|0\rangle$.
Critical coupling as a
function of $M$ and $N$ is shown in Fig.~8.
We see that the solution seems to be consistent with
our earlier results:  there is quick convergence as
$N$ increases and a logarithmic divergence as $M$ increases.

Thus, we believe that the {\em ansatz} introduced in
Eq.~(\ref{ansatz}) produces misleading results.
If one were to extend the {\em ansatz} to include more particles,
the convergent result for the critical coupling
would disappear.

As an example, let us examine some matrix elements of
$a_0$ for the truncation $M=3$ and $N=3$.
In this case, $S=20$ and we restrict our
attention to states with $P^{+}\leq3 (2 \pi)/d$
(numerical calculations are performed with all 20 states).
Arrange the basis states in the following
order:
\begin{equation}
\begin{array}{r@{}c@{}lc}
P^{+}&=&0 \;               &|0,0,0\rangle\\[10pt]
P^{+}&=&1\frac{2\pi}{d} \; &|1,0,0\rangle\\[10pt]
P^{+}&=&2\frac{2\pi}{d} \; &|0,1,0\rangle\\[5pt]
     & &                   &|2,0,0\rangle\\[10pt]
P^{+}&=&3\frac{2\pi}{d} \; &|0,0,1\rangle\\[5pt]
     & &                   &|1,1,0\rangle\\[5pt]
     & &                   &|3,0,0\rangle\\[10pt]
\multicolumn{4}{c}{\vdots}
\end{array}
\end{equation}
Given $g=2.8$ above the critical coupling
$g_{\rm critical} = 2.301$, we solve the constraint equation,
\begin{equation}
a_0=\left(\begin{array}{ccccc}
0&&&&\\
       &0&&&\\
&&\begin{array}{@{}cc@{}}
             0&-0.287\\
             -0.287&0
\end{array}&&\\&&&\begin{array}{@{}ccc@{}}
                       0&-0.281&0\\
                       -0.281&0&-0.458\\
                       0&-0.458&0
\end{array}&\\
                                   &&&&\ddots
\end{array}\right) \; .
\end{equation}
which leads to a $Z_2$ symmetric Hamiltonian.
Now, we lower the coupling to $g=1.8$.
The solution for the unbroken phase is now
\begin{equation}
a_0=\left(\begin{array}{ccccc}
0&&&&\\
       &0&&&\\
&&\begin{array}{@{}cc@{}}
             0&-0.318\\
             -0.318&0
\end{array}&&\\&&&\begin{array}{@{}ccc@{}}
                       0&-0.315&0\\
                       -0.315&0&-0.499\\
                       0&-0.499&0
\end{array}&\\
                                   &&&&\ddots
\end{array}\right) \; .
\end{equation}
In addition, we have two solutions for the broken phase.
There is a solution with $\langle 0|a_0|0\rangle>0$,
\begin{equation}
a_0=\left(\begin{array}{ccccc}
0.787&&&&\\
       &-0.246&&&\\
&&\begin{array}{@{}cc@{}}
             -0.251&-0.314\\
             -0.314&0.072
\end{array}&&\\&&&\begin{array}{@{}ccc@{}}
                       -0.270&-0.310&0.005\\
                       -0.310&0.070&-0.499\\
                       0.005&-0.499&-0.019
\end{array}&\\
                                   &&&&\ddots
\end{array}\right) \; ,
\end{equation}
%
%
%
along with the $\langle 0 |a_0|0\rangle <0$ solution,
\begin{equation}
a_0=\left(\begin{array}{ccccc}
-0.787&&&&\\
       &0.246&&&\\
&&\begin{array}{@{}cc@{}}
             0.251&-0.314\\
             -0.314&-0.072
\end{array}&&\\&&&\begin{array}{@{}ccc@{}}
                       0.270&-0.310&-0.005\\
                       -0.310&-0.070&-0.499\\
                       -0.005&-0.499&0.019
\end{array}&\\
                                   &&&&\ddots
\end{array}\right) \; .
\end{equation}
%
%

When we substitute the solutions for the broken phase of $a_0$ into the
Hamiltonian (\ref{hamiltonian}) we
get two Hamiltonians $H^+$ and $H^-$ corresponding to the two signs of
$\langle 0 | a_0 | 0 \rangle$ and the two branches of the curve in
Fig.~1.
This is the new paradigm for spontaneous symmetry breaking:
multiple vacua
are replaced by multiple Hamiltonians.
Picking the Hamiltonian defines the
theory in the same sense that picking the vacuum defines the theory in
the equal-time paradigm.
The two solutions for $a_0$ are related to each other
in a very specific way.  Let $\Pi$ be the unitary operator
associated with the $Z_2$ symmetry of the system;
$\Pi a_k \Pi^\dagger = -a_k$, $k \neq 0$.
We break up $a_0$ into an even part $\Pi a_0^E \Pi^\dagger = a_0^E$
and an odd part $\Pi a_0^O \Pi^\dagger = -a_0^O$.
The even part $a_0^E$ breaks the
$Z_2$ symmetry of the theory.
For $g<g_{\rm critical}$,
the three solutions of the constraint equation are:  $a_0^O$ corresponding to
the unbroken phase, $a_0^O+a_0^E$
corresponding to the
$\langle 0 | a_0 |0 \rangle >0$ solution, and
$a_0^O-a_0^E$ for the
$\langle 0 | a_0 |0 \rangle <0$ solution.
Thus, the two Hamiltonians are
\begin{equation}
H^+ = H\left(a_k, a_0 ^O + a_0^E\right)
\end{equation}
and
\begin{equation}
H^- = H\left(a_k, a_0^O - a^0_E\right)
\end{equation}
where H has the property
\begin{equation}
H\left(a_k, a_0\right) = H\left(-a_k, -a_0\right)
\end{equation}
and $a_k$ represents the nonzero modes.
Since $\Pi$ is a unitary operator,
if $|\Psi\rangle$ is an eigenvector of $H$ with
eigenvalue $E$ then $\Pi |\Psi\rangle$ is an
eigenvalue of $\Pi H \Pi^\dagger$  with eigenvalue $E$.
Since,
\begin{eqnarray}
\Pi H^- \Pi^\dagger &=&
\Pi H\left(a_k,a_0^O - a_0^E\right) \Pi^\dagger =
          H\left(-a_k ,-a_0^O-a_0^E\right) \nonumber\\
    &=& H\left(a_k,a_0^O +a_0^E\right) = H^+  \; ,
\end{eqnarray}
$H^+$ and $H^-$ have the same eigenvalues.

Using the $M=3$, $N=3$ case as an example, let
us examine the spectrum of $H$.
Removing the zero mode entirely, the
rescaled Hamiltonian, Eq.~(\ref{hamiltonian}) is
\begin{equation}
H=\left(\begin{array}{ccccc}
0&&&&\\
       &g&&&\\
&&\begin{array}{@{}cc@{}}
             \frac{g}{2}&0\\
             0&3+2 g
\end{array}&&\\&&&\begin{array}{@{}ccc@{}}
                       \frac{g}{3}&0&\sqrt{2}\\
                       0&3+\frac{3 g}{2}&0\\
                       \sqrt{2}&0&9+3 g
\end{array}&\\
                                   &&&&\ddots
\end{array}\right) \; .
\end{equation}
For large $g$ the eigenvalues are obviously: $0$, $g$, $g/2$, $2g$,
$g/3$, $3g/2$ and $3g$.
However as we decrease $g$ one of the last three
eigenvalues will be driven negative.
This signals the breakdown of the theory near
the critical coupling when the zero mode is not included.

Including the zero mode  fixes this problem.
Fig.~9 shows the spectrum for the three lowest nonzero
momentum sectors.
This spectrum illustrates several characteristics which
seem to hold generally (at least for truncations
 we have examined,  $N+M \leq 6$).
For the broken phase, the vacuum is the lowest energy
state, there are no level crossings as a function of $g$,
and the theory does not break down
in the vicinity of the critical point.
None of these are true for the spectrum with the
zero mode removed or for the unbroken phase
below the critical coupling.
The lowest eigenvalue has a minimum at or near the
critical point and the minimum appears to decrease
with the number of modes and particles.
A more precise statement will have to await a
fully renormalized treatment.
\section{Spectrum of the Field Operator}
\setcounter{equation}{0}
So far we have examined the zero mode and its effect on the
spectrum of the Hamiltonian.
How does the zero mode affect the field itself?
Since $\phi$ is a Hermitian operator it is an observable of
the system and one can measure $\phi$ for a given
state $|\alpha\rangle$.
The result is given by simple quantum mechanics.
Let us define the eigenvalues $\tilde{\phi}_i$ and eigenvectors
$|\chi_i\rangle$ of $\sqrt{4 \pi} \phi\,$:
\begin{equation}
 \sqrt{4 \pi} \phi \, |\chi_i\rangle =
\tilde{\phi}_i|\chi_i\rangle \;,
\;\;\;\; \langle \chi_i | \chi_j \rangle = \delta_{i,j}\; .
\end{equation}
The probability of obtaining $\tilde{\phi}_i$
as the result of a measurement of $\sqrt{4 \pi} \phi$
for the state $|\alpha\rangle$
is $\left|\langle \chi_i | \alpha \rangle \right|^2$.

In the limit of large $N$, the probability distribution
becomes continuous.  If we ignore the zero mode, the probability
of obtaining $\tilde{\phi}$ as the result of a measurement of
$\sqrt{4 \pi} \phi$ for the vacuum state is
\begin{equation}
P\left(\tilde{\phi}\right) = \frac{1}{\sqrt{2 \pi \tau}}\,
\exp\left(-\frac{\tilde{\phi}^2}{2 \tau}\right)\, d\tilde{\phi}
\label{gaussian}
\end{equation}
where $\tau = \sum_{k=1}^M 1/k$.  The probability distribution comes
from the
ground state wave function of the Harmonic oscillator
where we identify $\phi$ with the position operator.
This is just the Gaussian fluctuation of a free field.
Note that the width of the gaussian
diverges logarithmically in $M$.
When $N$ is finite, the distribution becomes discrete as
shown in Fig.~10.

In general, there are $N+1$
eigenvalues such that $\langle \chi_i | 0 \rangle \neq 0$,
independent of $M$.
Thus if we want to examine the spectrum of the field
operator
for the vacuum state, it is better to choose Fock space
truncations where $N$ is large.
With this in mind, we examine the $N=50$ and $M=1$ case as a
function of $g$ in Fig.~11.
Note that near the critical point, Fig.~11a, the  distribution
is approximately equal to the free field case shown in Fig.~10.  There
is no
symmetry breaking and the field is symmetric about zero.
As we move away from the critical point, Figs.~11b-d, the distribution
becomes increasingly narrow with a peak located at the VEV
of what would be the minimum of the symmetric double well
potential in the equal-time paradigm.
In addition, there is a small peak corresponding to minus the VEV.
In the language of the equal-time paradigm,
there is tunneling between the two minima of the potential.\\
\section{Discussion}
\setcounter{equation}{0}
In the context of $\left(\phi^4\right)_{1+1}$ on the light
front,
the vacuum of the full theory is always the perturbative
vacuum.  The zero mode, which satisfies an
operator valued constraint equation,
produces the long range physics
of the theory including spontaneous breaking of the
$Z_2$ symmetry.
We have found that the constraint equation can be solved
using a Tamm-Dancoff truncation of the Fock space.
Even for the one mode truncation, we find a critical
coupling consistent with the best equal time
calculations.  Increasing the Tamm-Dancoff truncation,
we find rapid convergence with the total number of particles
$N$.
This is to be contrasted with the equal-time
approach where an infinite number of particles
are required to produce a critical point.
In the weak coupling limit, the zero mode
ensures that light-front perturbation
theory agrees with equal-time perturbation
theory.  Above the critical coupling the zero mode develops
a contribution that breaks the $Z_2$ symmetry of the theory.
There are two such solutions to the constraint equation:
one with $\langle 0 |\phi|0\rangle >0$ and
one with $\langle 0|\phi|0\rangle <0$.
This, in turn, gives rise to two Hamiltonians
which have the same spectrum.
The spectrum has the expected behavior:
the Fock vacuum is the state of lowest energy and the
lowest eigenvalue has a minimum at the critical coupling.
Closer inspection of the field shows that
tunneling occurs between positive and negative
eigenvalues in the broken phase.

In this work, we apply a simple mass renormalization
to the symmetric ordered Hamiltonian.  This prescription
removes tadpoles from ordinary interaction terms and would
properly renormalize the theory if the zero mode were removed.
However, inclusion of the zero mode produces logarithmic
divergences in the constraint equation and in
the resulting Hamiltonian.
This can be seen from Figs.\ 6 and 8 where $g_{\rm critical}$
grows with the number of modes.  The number of modes
is equivalent to a large $P^+$ cutoff in a standard
discretized light-front quantization calculation.
We therefore believe that this behavior is a renormalization
effect related to the fact that the new interactions are not
normal ordered.
To a very good accuracy,
\begin{equation}
 g_{\rm critical} \approx
\left(3-\sqrt{3}\right) \sum_{m=1}^M \frac{1}{m} \; .
\end{equation}
This type of logarithmic growth indicates
that the system needs further renormalization.
The most direct approach would be to find
some non-perturbative method of normal ordering
the resulting Hamiltonian.
Another approach is to add a constant to $g$ and
multiply $\lambda$ by a constant so that logarithmic
divergences are removed.  In the one particle case, this
will work.  However, it is unclear if we can successfully
extend it to the more general case $N>1$.
Another possibility is to add more operators to the
Hamiltonian.  For instance, if we  add a term of the
form $\int dx^- \, dy^- \phi(x) \phi(y)$,
we can remove the logarithmic divergence in the
constraint equation.

For $\left(\phi^4\right)_{1+1}$ we believe that a
vanishing of the mass gap is associated with the critical
coupling~\cite{hari}.
In Fig.~7 we see that the gap between
the vacuum and the lowest energy excited state
is minimized at the critical coupling.
One could imagine that, in the limit of
large $M$, the gap between the vacuum and the first
excited state goes to zero at the critical coupling.
However, the first excited state is the one with the
largest longitudinal momentum and is always
at the ``edge'' of our Fock space truncation in $M$.
Thus, our Fock space truncation may not be particularly well suited
for an investigation of the vanishing of the mass gap.

We chose $\left(\phi^4\right)_{1+1}$ as a model
for its simplicity.
One problem with this choice is that the scalar field
$\phi$ is dimensionless.
Power counting arguments suggest that arbitrary powers of
$\phi$ are allowed in the Hamiltonian.
Consequently, the zero mode can be a very complicated
object for this model.
In contrast, for $\phi^4$ theory in (3+1)-dimensions the only
allowed local operators are  $\phi^n$, $n \leq 4$.
For physically interesting theories like QCD, the operators
that are allowed by power counting arguments are much more
restricted~\cite{GPSW}.
However, they still allow a very large set of operators and are therefore in
some
sense similar to the $\left(\phi^4\right)_{1+1}$ example.
Solution of the zero mode
problem for the model we chose may be
conceptually more difficult.

In theories like QCD we expect to have, in addition, zero modes
that are dynamical degrees of freedom.
A recent study of pure glue QCD in (1+1)-dimensions
shows that the zero mode of $A^+$ has this property~\cite{hcpauli,kalloniatis}.
Thus, in QCD, we expect to have dynamical
as well as nondynamical zero modes making
the problem quite complicated.
\section*{Acknowledgements}
The authors would like to acknowledge and thank  H-C. Pauli,
A. Kalloniatis, J. Hiller, G. McCarter, D. Robertson,
R. Perry and  A. Harindranath for many conversations and
useful comments.
This work was supported in part by grants from the
U. S. Department of Energy.
\newpage
\section*{Figure Captions}
\begin{description}

\item Figure 1. $f_0 = \sqrt {4 \pi} \langle 0 | \phi | 0 \rangle$
vs.\ $g = 24 \pi \mu^2/\lambda$  in the one mode case.
The solid curves are the critical curves obtained from numerical
solution
of Eq.~(\ref{recursion}) with $N=10$.
The dashed curve is the critical curve obtained from the
first-order $\delta$-expansion.

\item Figure 2. $\left|f_n\right|$ vs.\ $n$ and
$f_0$ for $g=1.2$ using Eq.~(\ref{recursion}).

\item Figure 3.  Critical curves for a larger range of $g$ in the one
mode case with $N=10$.
The curves shown are independent of $N$, $N \ge 10$.

\item Figure 4. The lowest three energy eigenvalues for the one mode
case as a function
of $g$ from the numerical solution of Eq.~(\ref{onehamiltonian})
with $N=10$.
The dashed lines are for the unbroken phase $f_0 = 0$ and
the solid lines are for the broken phase $f_0 \neq 0$.

\item Figure 5. $b_0= \sqrt {4 \pi} \langle 0 | \phi | 0 \rangle$ vs.\ $g$ for
the one particle case.
The critical curves are obtained from numerical solution
of Eqs.~(\ref{beforemixing}) and (\ref{mixing}) with $M=10$.
The vertical line represents $g_0$.
Note the presence of additional solutions
for $g$ small.

\item Figure 6.  $g_{\rm critical}$ vs.\ $M$ in the one particle case.
The solid curve is from Eq.~(\ref{loggc}) and the points are from
numerical solution of (\ref{beforemixing}) and (\ref{mixing}).
The dashed curve is $g_0$ from (\ref{g0}).

\item Figure 7.  Eigenvalues as a function of $g$
 in the one particle case using Eq.~(\ref{mham})
with  $M=10$.
The solid lines are for the broken phase and the
dashed lines are for the unbroken phase.  The
vertical line represents $g_0$.

\item Figure 8.  Critical coupling vs.\ $M$ and $N$.

\item Figure 9.  The spectrum for $M=3$, $N=3$ and (a)
$P^{+}=1 (2\pi)/d$, (b) $P^{+}=2 (2\pi)/d$,
and (c) $P^{+}=3 (2\pi)/d$.
The dashed line shows the spectrum with no zero mode.
The dotted line is the unbroken phase and the
solid line is the broken phase.

\item Figure 10.  Probability distribution of eigenvalues
of $\sqrt{4 \pi} \phi$ for the vacuum with
$M=1$, $N=10$, and no zero mode.  Also
shown is the infinite $N$ limit from Eq.~(\ref{gaussian}).

\item Figure 11.  Probability distribution of eigenvalues
of $\sqrt{4 \pi} \phi$ for the vacuum
with couplings (a) $g=1$, (b) $g=0$, (c) $g=-1$, and (d) $g=-2$.
$M=1$, $N=50$, and the positive VEV solution to the constraint
equation is used.

\end{description}
\end{document}